\documentclass[11pt]{article}

\usepackage{amsfonts}
\usepackage{epsf}
\usepackage{graphicx}

\begin{document}

\begin{center}
{\large\bf DIRECTED RANDOM WALK ON THE LATTICES\\ OF GENUS TWO}

\vspace*{0.5cm}
{A.V. Nazarenko}

\vspace*{0.5cm}
{Bogolyubov Institute for Theoretical Physics,\\
14-b, Metrologichna Str., Kiev 03680, Ukraine\\
nazarenko@bitp.kiev.ua}
\end{center}

\begin{abstract}
The object of the present investigation is an ensemble of self-avoiding and
directed graphs belonging to eight-branching Cayley tree (Bethe lattice) generated
by the Fucsian group of a Riemann surface of genus two and embedded in the Pincar\'e
unit disk. We consider two-parametric lattices and calculate the multifractal scaling
exponents for the moments of the graph lengths distribution as functions of these
parameters. We show the results of numerical and statistical computations, where the
latter are based on a random walk model.

\end{abstract}

\section{Introduction}

A number of stochastic models is taken up to describe the different structures on
underlying non-trivial geometrical carriers like random environments, percolation
clusters, fractals, etc. Restricting by physical systems and phenomena, we can note
turbulence~\cite{JKLPS,FP}, diffusion limited aggregation~\cite{SM}, polymers~\cite{Van,Hug},
the particle dynamics in a random magnetic field~\cite{CMW,CCFGM} as the examples.
The last decades it seems natural to describe these complicated and chaotic systems
by using the multifractality concept~\cite{Feder,Sor} giving us a spectrum of critical
exponents which can be connected with the results of renormalization group computations
(see Ref.~\cite{BJ}).

We can also find a great number of geometrical objects, chaotic at first
sight, within the Lobachevsky geometry. Indeed, even limiting by dimension two, there is
a possibility to construct the infinite number of hyperbolic lattices, existence of which
is not allowed in a flat space~\cite{DNF}. This fact inspires us in the present paper to
turn to Lobachevsky geometry and to demonstrate that the structures, well studied in
mathematics, might pretend to describe and generate some kind of disorder, like a random
walk, analytically. Similar idea to consider a directed random walk on a Cayley tree in
hyperbolic space has been already exploited in Refs.~\cite{VN,CNV}. We also note that the
models of billiards intensively use the hyperbolic Riemann surfaces as geometrical
carriers~\cite{Gutz,N}.

In our paper, we shall watch a thread of walker moving along the eight-branching
and rooted Cayley tree constructed as a dual lattice to the hyperbolic octagonal lattice
tiling the Poincar\'e disk under the action of Fucsian group. The octagonal lattice cells,
whose opposite sides are identified, correspond topologically to the two-holed torus,
geometry of which is defined here by one independent module while other two complex moduli
are constrained. (Note that the Riemann surfaces of genus $g$ are characterized by $6g-6$
real parameters.) We characterize a walk by spectrum of lengths which result from summation of
hyperbolic distances between tree sites visited and root point. Introducing a partition
function and the moments of distribution function, we would like to investigate an ensemble
of these lengths within the multifractality concept.

Thus, the model under consideration is deterministic by construction, and partition function
is actually the truncated series over a symmetry group. To calculate this kind of series, we
apply both numerical and statistical methods, where the former is needed to exhibit the basic
properties of the model, which we aim to reproduce by means of the latter approach
called for purely analytical description. Within the framework of statistical method,
we elaborate a scheme of length spectrum computation and reduce the partition function
to the Markov multiplicative chain in a random walk approximation.

\section{Space and Symmetries}

We will deal with a particular case of a two-dimensional Lobachevsky space (with
the Gaussian curvature $K=-1$), namely, a model of the open Poincar\'e disk
centered at the origin,
\begin{equation}
\mathbb{D}=\left\{(x,y)\in\mathbb{R}^2|x^2+y^2<1\right\},
\end{equation}
endowed with the metric:
\begin{equation}\label{Poin}
ds^2=4\frac{dx^2+dy^2}{(1-x^2-y^2)^2}.
\end{equation}

Mainly, we are interested in the symmetries (isometries) of the metric (\ref{Poin}),
which we will use for construction of non-trivial structures on $\mathbb{D}$.
However, an application of this metric may not be limited by symmetry production only.
In Appendix~A, we also develop heuristic ideas how to relate the curved space to
the Euclidean two-dimensional space with potential field. Such a correspondence,
in our mind, can follow from Maupertuis variational principle, constituting an
equivalence between particle trajectories (written down below) in these spaces.

Solution to the geodesic equation in $(\mathbb{D},ds^2)$ is the functions:
\begin{eqnarray}\label{xy}
x(s)&=&\frac{\cos{\phi}\cosh{s}+R\sin{\phi}\sinh{s}}{\sqrt{1+R^2}\cosh{s}+R},
\nonumber\\
y(s)&=&\frac{R\cos{\phi}\sinh{s}-\sin{\phi}\cosh{s}}{\sqrt{1+R^2}\cosh{s}+R},
\end{eqnarray}
which are defined in the interval $s\in(-\infty,+\infty)$ and describe an arc
inside $\mathbb{D}$ with the radius $R$ and the center at the point
$x_0=\sqrt{1+R^2}\cos{\phi}$, $y_0=\sqrt{1+R^2}\sin{\phi}$, lying beyond the
unit disk.

Now let us introduce a structure on the Poincar\'e disk. Here we aim to consider an
example of lattice resulting from operation of a particular Fuchsian group $\Gamma$
(stating ``periodic law'') on the (complexified) Poincar\'e disk $\mathbb{D}$ \cite{DNF}.
The associated quotient space $\mathbb{D}/\Gamma$ is assumed to be a Riemann surface of
genus $g=2$, whose fundamental domain ${\cal F}$ is hyperbolic octagon in $\mathbb{D}$,
which can be viewed as the result of gluing two tori together.

Thus we should first construct the fundamental octagon ${\cal F}$
and the corresponding Fuchsian group $\Gamma$ by means of which we tile $\mathbb{D}$ by
the octagonal lattice. Note that an algorithm connecting ${\cal F}$ for $g=2$
with Fuchsian group $\Gamma$ and vice versa is described in Ref.~\cite{ABCKS}.

The Riemann surface of genus $g=2$ is completely determined by a set of three complex
moduli corresponding to three corners of the fundamental domain ${\cal F}$ (in complex
plane). The fourth corner is then found to adjust the area of the fundamental domain to
be $2\pi(2g-2)=4\pi$, according to the Gauss-Bonnet theorem~\cite{DNF}. Applying inversions
across the origin to these four corners yield the remaining four corners of the octagon.

For the sake of simplicity, we shall focus on symmetric octagon with the {\it one} arbitrary
module, namely, one pair of the length and angle variables denoted as $(a,\alpha)$. Such an
octagon (sketched in Fig.~1) can be obtained by deformation of the regular hyperbolic
octagon (with $a=2^{-1/4}$, $\alpha=\pi/4$), well studied in the context of the chaology
(see, for example, Refs.~\cite{Gutz,ABCKS} and references therein). Introducing the complex
variable $z=x+iy\in\mathbb{C}$ in order to develop the further analysis in usual manner, this
construction is as follows.

Let us assume that the corners of the octagon are at the points $a\exp{(ik\pi/2)}$,
$b\exp{i(\alpha+k\pi/2)}$, where $0<a,b<1$, $k=\overline{0,3}$. At this stage, parameter
$b$ is unknown and should be found as the function of $(a,\alpha)$. Connecting the points
by geodesics, we shall require the sum of inner angles (by vertices) of the octagon to be
equal $2\pi$ in accordance with the Gauss-Bonnet theorem. This requirement allows us to
determine $b$ and the parameters of geodesics constituting sides of octagon. It is obvious
that $b=a=2^{-1/4}$ for the regular octagon (the fixed value of $a$ is required because
the homothety property is not valid in the space with $K\not=0$).

\begin{figure}
\begin{center}
\includegraphics[width=4.5cm]{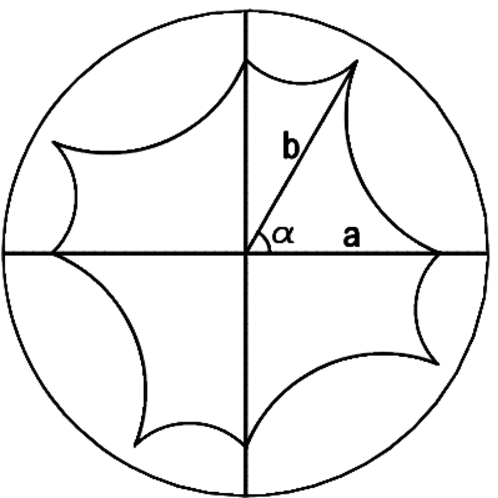}\qquad
\includegraphics[width=4.6cm]{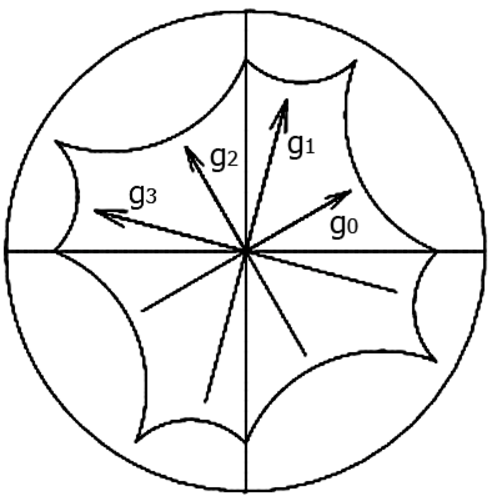}
\end{center}
\vspace*{-3mm}
\caption{\small Symmetric octagon with $a=0.8$, $\alpha=\pi/3$ and generators $g_k$ of
Fuchsian group.}
\end{figure}

In the case at a hand, the boundary of octagon, $\partial{\cal F}$, is formed by geodesics
of two kinds (labeled by ``$\pm$'' below). These geodesics are completely determined by the
radii $R_\pm$ and the angles $\phi_\pm+k\pi/2$ ($k=\overline{0,3}$), defining the
positions of the circle centers. Satisfying the conditions imposed above and collected in
manifest form in Appendix~B, we find that
\begin{equation}\label{R-phi}
R_\pm=\frac{1}{2a}\sqrt{T^2_\pm+(1-a^2)^2},\qquad
\phi_\pm=\arctan{\left[\left(\frac{T_\pm}{1+a^2}\right)^{\pm1}\right]},
\end{equation}
where
\begin{equation}
T_\pm=a^2\pm\tan{\left(\alpha-\frac{\pi}{4}\right)}.
\end{equation}

In these formulas it is supposed that $0<\phi_+<\alpha<\phi_-<\pi/2$. Moreover,
introducing an angle $\beta$ by vertices $a\exp{(ik\pi/2)}$ (the angle by
vertices $b\exp{i(\alpha+k\pi/2)}$ is then equal to $\pi/2-\beta$),
\begin{eqnarray}
\tan{\beta}&=&\frac{1-\sqrt{1+R^2_+}\sqrt{1+R^2_-}\cos{(\phi_--\phi_+)}}
{1-\sqrt{1+R^2_+}\sqrt{1+R^2_-}\sin{(\phi_--\phi_+)}}\nonumber\\
&=&(1-a^2)\frac{2a^2\cos^2\left(\alpha-\frac{\pi}{4}\right)}
{2a^2\cos^2\left(\alpha-\frac{\pi}{4}\right)-1},
\end{eqnarray}
we should control the condition $0<\beta<\pi/2$, defining the region of accessibility
of parameters $(a,\alpha)$:
\begin{equation}
\frac{1}{\sqrt{2}\cos{\left(\alpha-\frac{\pi}{4}\right)}}<a<1,
\end{equation}
which is sketched in Fig.~2.

\begin{figure}
\begin{center}
\includegraphics[width=8cm]{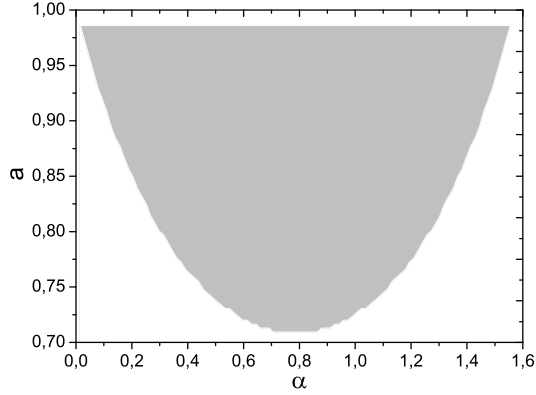}
\end{center}
\vspace*{-10mm}
\caption{\small The permission domain of module $(a,\alpha)$.}
\end{figure}

To complete the octagon description, the parameter $b$, pointed out in Fig.~1, can be
calculated as
\begin{eqnarray}\label{b}
b&=&\cos{(\alpha-\phi_+)}\left[\sqrt{1+R^2_+}-\sqrt{R^2_+-\tan^2{(\alpha-\phi_+)}}\right]
\nonumber\\
&=&\frac{1}{\sqrt{2}a\cos{(\alpha-\pi/4)}}.
\end{eqnarray}

Note that the manifest dependence of the octagon parameters on moduli is necessary
in the different problems where geometry of the system (carrier geometry) is not fixed.
For instance, $(a,\alpha)$ would be the (dynamical) variables of evolution parameter in
topological gravity describing an universe of genus $g=2$; it is able to average
over $(a,\alpha)$ in statistical physics, etc.

The Fuchsian group $\Gamma$ of a Riemann surface of genus two is generated by four
generators $g_0$, $g_1$, $g_2$, $g_3$ and their inverses. Let generators $g_k$,
$g^{-1}_k$, acting freely as isometries of Poincar\'e disk, map geodesic boundary
segments of octagon ${\cal F}$ onto each other, thereby identifying opposite edges
(see Fig.~1, right panel). This definition leads to the group relation:
\begin{equation}
g_0 g^{-1}_1 g_2 g^{-1}_3 g^{-1}_0 g_1 g^{-1}_2 g_3={\rm id}.
\end{equation}

Generally, the group of orientation-preserving isometries of $(\mathbb{D},ds^2)$
is presented by $PSU(1,1)=SU(1,1)/\{\pm1\}$ (see textbooks like Ref.~\cite{DNF}), where
\begin{equation}
SU(1,1)=\left\{\left.
\left(\begin{array}{cc}
u&v\\
\overline{v}&\overline{u}
\end{array}\right)\right| u,v\in\mathbb{C},
|u|^2-|v|^2=1
\right\}.
\end{equation}
The group of isometries acts via fractional linear transformations:
\begin{equation}
z\mapsto \gamma[z]=\frac{uz+v}{\overline{v}z+\overline{u}},\quad
\gamma\in PSU(1,1),\quad
z\in\mathbb{D}=\{z\in\mathbb{C}||z|<1\}.
\end{equation}
Then the Fuchsian group $\Gamma$ is a discrete group such that
$\Gamma\subset PSU(1,1)$. One also adds the hyperbolicity condition,
$|{\rm Tr}~\gamma|>2$ for all $\gamma\in\Gamma$. Constructing the
corresponding Fuchsian group for a given octagon, it is convenient to
express the generators of $\Gamma$ by half turns as follows.

Let $p_k$ be the mid-point of $k$-th side, $k=\overline{0,3}$. Since the
opposite sides of octagon have the same lengths by construction, the
generators are then written as $g_k=H(p_k)$ (see Ref.~\cite{ABCKS}), where
\begin{equation}\label{mat1}
H(p)=\frac{-1}{1-|p|^2}\left(
\begin{array}{cc}
	1+|p|^2& 2p\\
	2\overline{p}& 1+|p|^2
\end{array}
\right)\in PSU(1,1).
\end{equation}
Bar over expression, $\overline{p}$, means the complex conjugation.

The operation of matrices $H(p)$ consists in the half turn (rotation
with angle $\pi$) of geodesic segment around the origin $z=0$ and the half turn
around point $p$.

In our model, all $p_k$ are parametrized by $p_\pm\in\mathbb{D}\subset\mathbb{C}$:
$p_0=p_+$, $p_1=p_-$, $p_2=p_+{\rm e}^{i\pi/2}$, and $p_3=p_-{\rm e}^{i\pi/2}$.
At this time, $p_\pm$ are functions of two parameters $(a,\alpha)$ and written down
below. In principal, we could first take $p_k$ as ``input'' and, using then these
points, determine the octagon parameters.

To find $p_k$ analytically, set
\begin{equation}
\omega_\pm=\frac{b{\rm e}^{i\alpha}(1-a^2)+a{\rm e}^{i\pi(1\mp1)/4}(1-b^2)}{1-a^2b^2},
\end{equation}
then one has that
\begin{equation}
p_\pm=\frac{\omega_\pm}{1+\sqrt{1-|\omega_\pm|^2}}.
\end{equation}

The generators $g_k$ can be directly expressed in the terms of auxiliary variables
$\omega_\pm$. Therefore, $g_k=M(\omega_k)$, where
\begin{equation}\label{mat2}
M(\omega)=\frac{-1}{\sqrt{1-|\omega|^2}}\left(
\begin{array}{cc}
	1& \omega\\
	\overline{\omega}& 1
\end{array}
\right)\in PSU(1,1),
\end{equation}
and $\omega_0=\omega_+$, $\omega_1=\omega_-$, $\omega_2=\omega_+{\rm e}^{i\pi/2}$,
$\omega_3=\omega_-{\rm e}^{i\pi/2}$.

Now it is possible to build the octagonal lattice with a given fundamental octagon.
By the construction, action of the group elements $\gamma\in\Gamma$ on ${\cal F}$
produces the daughter cells tiling Poincar\'e disk,
$$
\mathbb{D}=\bigcup\limits_{\gamma\in\Gamma}\gamma{\cal F}.
$$

\begin{figure}
\begin{center}
\includegraphics[width=4.5cm]{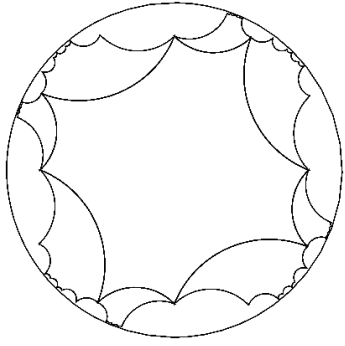}
\end{center}
\vspace*{-3mm}
\caption{\small The fundamental octagon with $a=0.8$, $\alpha=\pi/3$ and the daughter cells.}
\end{figure}
The first eight cells $g^{\pm1}_k{\cal F}$ ($k=\overline{0,3}$) are shown in Fig.~3.
We see that $g^{\pm1}_k{\cal F}\cap{\cal F}\subset\partial{\cal F}$ as must be.
Of course, this property is preserved for all neighboring cells (not only for
fundamental polygon ${\cal F}$).

\section{The Directed Random Walk}

In this Section, we concern with the directed random (self-avoiding) walk (DRW) on a
8-branching Cayley tree isometrically embedded in Poincar\'e disk. The tree under
consideration is formed by the graph connecting the centers of the neighboring
octagons (see Fig.~4) and may be physically interpreted as a model of directed polymers.
\begin{figure}
\begin{center}
\includegraphics[width=5cm]{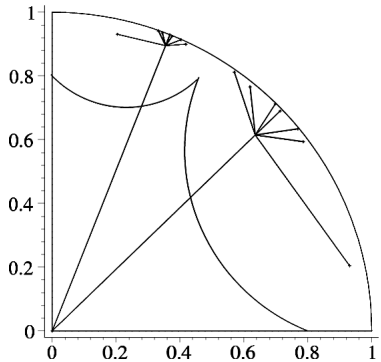}
\end{center}
\vspace*{-3mm}
\caption{\small The quarter of fundamental octagon with $a=0.8$, $\alpha=\pi/3$, and the
Cayley tree sites of first and second generations (connected schematically by straight lines
but not geodesics).}
\end{figure}

We start from the general expression for partition function of DRW in disordered
environment on two-dimensional lattice:
\begin{equation}
{\cal Z}_N(q)=\sum\limits_{\{z\}}\prod\limits_{t=1}^N p(z_{t-1}|z_t)
\exp{\left[q V(z_t,t)\right]},
\end{equation}
where $N$ is the number of generations; $z_t$ is a point of $t$-th generation of random
trajectory; $V$ is a random potential of environment; $p$ is a transition probability;
parameter $q$ can be mapped into an inverse temperature.

In our model, we consider a deterministic motion on graphs of $N$-th
generation, sites of which correspond to a set of numbers
$\{i_1,i_2,\ldots i_N\}$, where $1\leq i_t\leq 8$, $1\leq t\leq N$.
We also assume that
\begin{equation}
z_t=\gamma_{i_1}\gamma_{i_2}\ldots\gamma_{i_t}[0],\qquad
V(z_t,t)=d(0,z_t)/N,
\end{equation}
\begin{equation}
p(z_{t-1}|z_t)\to p(i_{t-1}|i_t)=\left\{
\begin{array}{cc}
1, &|i_t-i_{t-1}|\not=4\\
0,	&|i_t-i_{t-1}|=4
\end{array}
\right.,
\end{equation}
where $\{\gamma_i|i=\overline{1,8}\}=\{g_0,g_1,g_2,g_3,g^{-1}_0,g^{-1}_1,g^{-1}_2,g^{-1}_3\}$.
By construction, points $z_t$ and $z_{t-1}$ are the centers of neighbouring cells connected by
the tree branches.
Definition of $p(i_{t-1}|i_t)$ (consistent with definition of generators $\gamma_i$,
$i=\overline{1,8}$) excludes a return to the visited sites. Hyperbolic distance $d(z,w)$ on
$(\mathbb{D},ds^2)$ is determined by
\begin{equation}
\cosh{d(z,w)}=1+\frac{2|z-w|^2}{(1-|z|^2)(1-|w|^2)}.
\end{equation}

In particular case, one has
\begin{equation}
d(0,z)=\ln{\frac{1+|z|}{1-|z|}}.
\end{equation}

In this way, we come to expression for partition function of our model:
\begin{equation}\label{partfunc}
{\cal Z}_N(q)=\sum\limits_{i_1=1}^8\ldots\sum\limits_{i_N=1}^8
\left(\prod\limits_{t=1}^N p(i_{t-1}|i_t)\right)
\exp{\left[q\frac{1}{N}\sum\limits_{t=1}^N \rho_{i_1,i_2\ldots i_t}\right]},
\end{equation}
where $\rho_{i_1,i_2\ldots i_t}\equiv d(0,\gamma_{i_1}\gamma_{i_2}\ldots\gamma_{i_t}[0])$
is the hyperbolic (geodesic) distance between the point of $t$-th generation and the origin $O$.

The partition function defined is a sum of Boltzmann weights that depend on an ``action 
integral'' linear in the hyperbolic distance from a given root point (the origin $O$), i.e. the
length of trajectory:
\begin{equation}
L(i_1,i_2,\ldots,i_N)\equiv\sum\limits_{t=1}^N \rho_{i_1,i_2\ldots i_t}.
\end{equation}

It is worthy to note that ${\cal Z}_N(q)$ is also a function of arbitrary
module $(a,\alpha)$, used for obtaining the generators of the group $\Gamma$. To explore the
properties of a model described by ${\cal Z}_N(q|a,\alpha)$, we shall fix the lattice parameters.

Since the all trajectories are self-avoiding and uniquely determined, the number of graphs
$Z_N(0)$ is equal to the number of end-points, $v(N)=8\times7^{N-1}$, and the partition function
(\ref{partfunc}) can be re-written as
\begin{equation}\label{parf}
{\cal Z}_N(q)=\sum\limits^{v(N)}_{i=1}{\rm e}^{q L_i(N)/N},
\end{equation}
where $L_i(N)$ is the integrated hyperbolic length of $i$-th graph of $N$-th generation.

Thus, the main goal of this work is to evaluate the series (\ref{parf}) dependent on
length spectrum $\{L_i(N)\}$. To resolve this problem, we shall first perform numerical
investigations to establish basic properties of such a system. Secondly, we shall attempt
to calculate the characteristics analytically with the use of statistical approach.

An example of length spectrum obtained numerically is presented in Fig.~5, where the
absolute value of the number of events depends on particular choice of the width of shell.
We see that the distribution of lengths $L_i$ approaches to the Gaussian distribution,
\begin{equation}
\sim\exp{\left(-\frac{(L-{\bar L}_N)^2}{2\sigma^2_N}\right)},
\end{equation}
with parameters defined by formulas
\begin{equation}
{\bar L}_N=\frac{1}{v(N)}\sum\limits_{i=1}^{v(N)}L_i(N),\quad
\sigma^2_N=\frac{1}{v(N)}\sum\limits_{i=1}^{v(N)}(L_i(N)-{\bar L}_N)^2,
\end{equation}
resulting to ${\bar L}_5\approx41.97$, $\sigma^2_5\approx44.5657$.
 
Although the process of such a convergence is slow, it means that the structure generated
by the discrete group $\Gamma$ satisfactory describes a disordered environment that justifies
a random walk treatment. Also note that although Fig.~5 was built for the fixed parameters
of the lattice, the system reveals the  same tendency for any admissible pair $(a,\alpha)$. 

\begin{figure}
\begin{center}
\includegraphics[width=9cm]{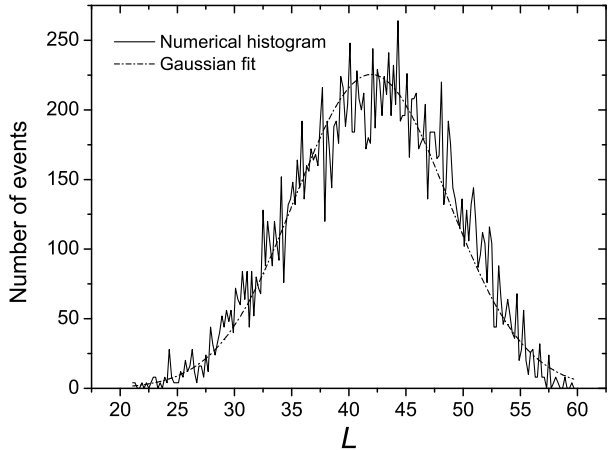}
\end{center}
\vspace*{-10mm}
\caption{\small The number of events as the function of the hyperbolic length $L(N=5)$
and the Gaussian fit for lattice with $a=0.8$, $\alpha=\pi/3.$}
\end{figure}

We are giving the further analysis of the model within the framework of the concept of
multifractality consisting in a scale dependence of critical exponents~\cite{Feder}.
To do this, let us introduce the multifractal moments of the order $q$ defined
as follows
\begin{equation}\label{moms}
\mathfrak{m}^{(q)}_N=\frac{{\cal Z}_N(q)}{{\cal Z}^q_N(1)}.
\end{equation}

Then the scaling exponent, sometimes called as mass spectrum, is
\begin{equation}
\tau_N(q)=\frac{2}{\ln{v(N)}}\ln{\mathfrak{m}^{(q)}_N},
\end{equation}
where prefactor 2 corresponds to dimension of underlying geometrical carrier (space)
and $v(N)$ coincides with the number of ``boxes'' tilling structure under consideration.

Exponent $\tau_N(q)$ is self-averaging and therefore it is reasonable to introduce
\begin{equation}
\tau(q)=\lim\limits_{N\to\infty}\tau_N(q),
\end{equation}
defining the spectrum of fractal dimensions $D_q=\tau(q)/(1-q)$.

The result of our numerical calculations (limited to $N\leq5$) is shown in Fig.~6.
We can conclude that the multifractal behavior takes place indeed, that is reflected
in non-linear dependence $\tau$ on $q$.
\begin{figure}
\begin{center}
\includegraphics[width=8.5cm]{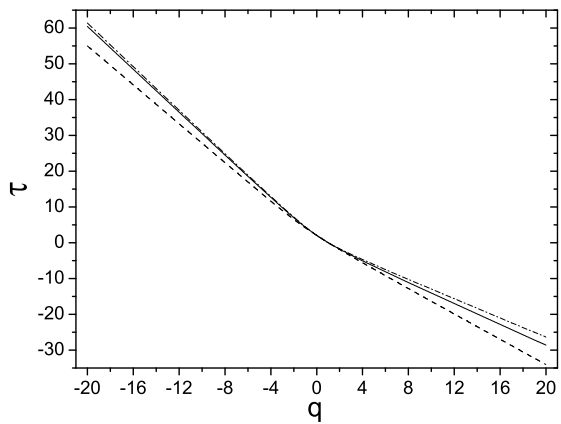}
\end{center}
\vspace*{-10mm}
\caption{\small Multifractal behavior of the exponent $\tau(q)$ for $N=5$. The solid, dashed,
and dashed-dotted curves correspond to the lattices with $(a=0.8,\alpha=\pi/3)$,
$(a=2^{-1/4},\alpha=\pi/4)$, and $(a=0.9,\alpha=\pi/8)$, respectively.}
\end{figure}

Besides the finding dependence $\tau(q)$, there is another equivalent way to  describe
the multifractality. It is well known that the set of exponents governing of multifractal
moments of type (\ref{moms}) is related to the spectrum of singularities $f(\alpha)$
of fractal measure,\cite{Halsey} called also the spectral function, which is given
by the Legendre transform:
\begin{equation}
f(\alpha)=q\alpha+\tau(q),\qquad
\alpha(q)=-\frac{d\tau(q)}{dq},
\end{equation}
where $\alpha$ is known as Lipschitz--H${\ddot {\rm o}}$lder exponent\footnote{We preserve
conventional symbol for this exponent~\cite{Feder}. It will always be clear from context whether
$\alpha$ is an angle determining the geometry of octagon or the singularity strength.}. In other
words, $f(\alpha)$ denotes the dimension of the subset characterized by the singularity strength
$\alpha$. In the terms of thermodynamics, $\alpha(q)$ coincides with an internal energy $e$,
while $f(\alpha)$ is interpreted as an entropy $s(e)$ \cite{JKP}.

The general properties of $f(\alpha)$ are as follows: it is positive on an
interval $[\alpha_{\rm min},\alpha_{\rm max}]$, where
\begin{equation}
\alpha_{\rm min}=\lim\limits_{q\to+\infty}\tau(q)/(1-q),\qquad
\alpha_{\rm max}=\lim\limits_{q\to-\infty}\tau(q)/(1-q),
\end{equation}
and the maximum value of the spectral function gives us the (fractal) dimension
of the underlying structure.

Spectral function $f(\alpha)$ obtained for $N=5$ is given in Fig.~7. In order to
compare curves in Fig.~7, we appeal to an information entropy $S$~\cite{Feder},
defined as
\begin{equation}
S=\alpha_S=f(\alpha_S),\qquad
\left.\frac{df(\alpha)}{d\alpha}\right|_{\alpha_S}=1.
\end{equation}
The value of $\alpha_S$ and $f(\alpha_S)$ can be directly calculated at $q=1$.

We find that $S(a=0.9,\alpha=\pi/8)\approx1.783<S(a=0.8,\alpha=\pi/3)\approx1.865<
S(a=2^{-1/4},\alpha=\pi/4)\approx1.95$. This relation has a simple treatment:
i) due to a special symmetry, the regular octagon gives us a highest entropy which
cannot exceed 2 in our model; ii) deviation from the regular symmetry leads to
decreasing the entropy more and more, what says about low probable configurations. 

\begin{figure}
\begin{center}
\includegraphics[width=8.5cm]{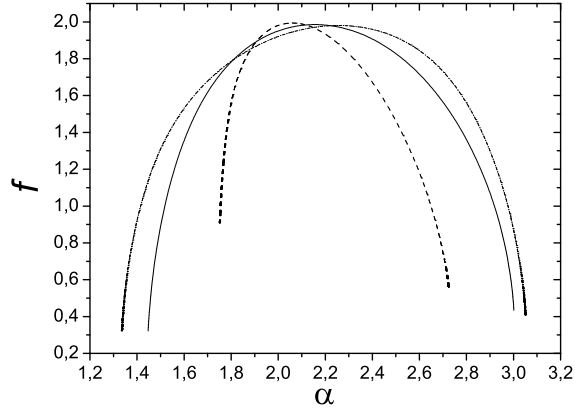}
\end{center}
\vspace*{-10mm}
\caption{The spectrum of singularities $f(\alpha)$ for $N=5$. The solid, the dashed,
and the dashed-dotted curves correspond to the lattices with $(a=0.8,\alpha=\pi/3)$,
$(a=2^{-1/4},\alpha=\pi/4)$, and $(a=0.9,\alpha=\pi/8)$, respectively.}
\end{figure}

Since the problem of random walk is mathematically the problem of addition of independent
random variables, the ultimate behavior of random walkers can be deduced from a central
limit theorem (CLT). Thus, another approach, applying CLT, can be also developed.

First, we need to reduce an evolution to the Markov multiplicative process. In order to
achieve this, let us use the triangle rule in two-dimensional hyperbolic space:
\begin{equation}
\cosh{d(0,z_t)}=\cosh{d(0,z_{t-1})}\cosh{d(z_t,z_{t-1})}
-\sinh{d(0,z_{t-1})}\sinh{d(z_t,z_{t-1})}\cos{\theta_{t,t-1}},
\end{equation}
where $\theta_{t,t-1}$ is angle opposite to side $(0,z_t)$.

Due to identities
\begin{equation}
d(0,\gamma^n[0])=nd(0,\gamma[0]),\qquad
d(\tilde\gamma[0],\tilde\gamma\gamma[0])=d(0,\gamma[0]),
\end{equation}
which are valid for matrices (\ref{mat1}), (\ref{mat2}), we get that
$d(z_t,z_{t-1})=d(0,\gamma_{i_t}[0])$ and
\begin{equation}
\cosh{d(0,z_t)}=\cosh{d(0,z_{t-1})}\cosh{\ell_{i_t}}
-\sinh{d(0,z_{t-1})}\sinh{\ell_{i_t}}\cos{\theta_{t,t-1}},
\end{equation}
where, accounting for lattice symmetry, $\ell_i=\delta_i^+\ell_++\delta_i^-\ell_-$, 
\begin{eqnarray}
&&\ell_+=d(0,\gamma_{2k-1}[0]),\quad \ell_-=d(0,\gamma_{2k}[0]),\quad
k=\overline{1,4},\\
&&\delta_i^+=\sum\limits_{k=1}^4\delta_{i,2k-1},\quad
\delta_i^-=\sum\limits_{k=1}^4\delta_{i,2k}.\label{delta}
\end{eqnarray}

For relatively large $d(0,z_t)$ and $d(0,z_{t-1})$, one has
\begin{equation}
d(0,z_t)=d(0,z_{t-1})+\ln{(\cosh{\ell_{i_t}}-\sinh{\ell_{i_t}}\cos{\theta_{t,t-1}})}.
\end{equation}

Now it is useful to introduce
\begin{equation}
\xi_{i,j}=d(0,\gamma_i\gamma_j[0])-\frac{\ell_i+\ell_j}{2},
\qquad i,j=\overline{1,8},
\end{equation}
describing a logarithmic contribution for the second generation.

Then, neglecting fluctuations of $\theta_{t,t-1}$ in higher generations,
we obtain an estimation for trajectory length:
\begin{equation}\label{len}
L(i_1,i_2,\ldots,i_N)\simeq N\ell_{i_1}+\sum\limits_{t=2}^N(N+1-t)\xi_{i_{t-1},i_t}
\quad {\rm for} \quad p(i_{t-1}|i_t)\not=0.
\end{equation}
Note immediately that this formula is exact for $i_1=i_2=\ldots=i_N$ when $\theta_{t,t-1}=\pi$.

In principal, it is also possible to introduce quantities 
$\xi^{(n)}_{i_1,i_2,\dots,i_n}\sim d(0,\gamma_{i_1}\gamma_{i_2}\dots\gamma_{i_n}[0])$ playing a
role of higher-order correlations and asymmetric by indices in contrast with symmetric matrix
$\xi_{i,j}=\xi_{j,i}$. Moreover, matrices $\xi^{(n)}$ are able to account more precisely for
values of angles $\theta_{t,t-1}$ or, in another word, fluctuations.

Using Eq.~(\ref{len}), we can easily give the theoretical estimation for the lengths spectrum
of $N$-th generation
\begin{eqnarray}
L^{\rm th}_{\rm min}(N)&=&N{\rm min}(\ell_+,\ell_-)+\frac{N(N-1)}{2}{\rm min}(\xi_{i,j}),
\quad {\rm for} \quad p(i|j)\not=0,\nonumber\\
{\bar L}^{\rm th}_N&=&N\frac{\ell_++\ell_-}{2}+\frac{N(N-1)}{2}\bar\xi,
\qquad \bar\xi=\frac{1}{56}\sum\limits_{i,j=1}^8p(i|j)\xi_{i,j},\label{lengths}\\
L^{\rm th}_{\rm max}(N)&=&N{\rm max}(\ell_+,\ell_-)+\frac{N(N-1)}{2}{\rm max}(\xi_{i,j})
\nonumber\\
&=&\frac{N(N+1)}{2}{\rm max}(\ell_+,\ell_-).\nonumber
\end{eqnarray}

Substituting (\ref{len}) in (\ref{partfunc}), we arrive at the expression for
partition function as inhomogeneous Markov multiplicative chain:
\begin{equation}\label{parfM}
{\cal Z}_N(q)\simeq\sum\limits_{i_1=1}^8\ldots\sum\limits_{i_N=1}^8
\exp{(q\ell_{i_1})}\prod\limits_{t=2}^N {\cal P}^t(i_{t-1}|i_t),
\end{equation}
where exponent function plays a role of initial condition, and
\begin{equation}
{\cal P}^t(i_{t-1}|i_t)=p(i_{t-1}|i_t)\exp{\left[q\frac{N+1-t}{N}\xi_{i_{t-1},i_t}\right]}
\end{equation}
is transition weight dependent on $t$. Properties of Eq.~(\ref{parfM}) are
discussed in details in Appendix~C.

Thus, we would like to note that the length spectrum $\{L_i(N)\}$ and the
corresponding partition function for any $N$ are completely determined by minimal
lengths $\ell_\pm$ and matrix $\xi_{i,j}$ in a given approximation.

Now, arranging the lengths of graphs in ascending order $L_1\leq L_2\leq\ldots\leq L_s$
($s<v(N)$), we can re-write the partition function (\ref{parf}) as
\begin{equation}\label{parfn}
{\cal Z}_N(q)=v(N)\sum\limits^{s}_{i=1}W_i{\rm e}^{q l_i},\quad
\sum\limits^{s}_{i=1}W_i=1,
\end{equation}
where $l_i=L_i/N$ and $W_i$ denotes the weight of the orbit of length $L_i$.

Hereafter, coefficients $W_i$ are assumed to be randomly distributed. It means
that the value of $W_i$ is independent of $W_k$ for $i\not=k$, allowing the application
of the random walk model.

Next, assuming that the CLT is valid at $N\to\infty$ for our system, let us model
a distribution of the lengths $l_i$ of $N$-th generation graphs by the Gaussian
distribution:
\begin{equation}\label{gauss}
W_N(l)=A_N\exp{\left(-\frac{(l-N{\bar l})^2}{2Ns^2}\right)},
\end{equation}
where $N{\bar l}={\bar L}_N/N$, $Ns^2=\sigma^2_N/N^2$ when $N\to\infty$; $A_N$ is
a normalization constant. We can theoretically evaluate ${\bar l}$ and $s^2$ by the
use of Eqs.~(\ref{len}), (\ref{lengths}).

We have already seen in Fig.~5 that the Gaussian approximation is in good agreement 
with numerical data. So, it is believed that an application of Gaussian distribution
may simplify significantly the calculations of the series over group $\Gamma$.

Applying the approximate distribution (\ref{gauss}), we can replace the sums (\ref{parfn})
with integrals at large $N$:
\begin{equation}\label{parf1}
{\cal Z}_N(q)\approx v(N)\int\limits^{\ell_{\rm max}(N)}_{\ell_{\rm min}(N)}W_N(l){\rm e}^{q l} dl,
\quad
\int\limits^{\ell_{\rm max}(N)}_{\ell_{\rm min}(N)}W_N(l)dl=1,
\end{equation}
where $\ell_{\rm min}(N)=L_{\rm min}(N)/N$ and $\ell_{\rm max}(N)=L_{\rm max}(N)/N$ are
introduced in order to cut long Gaussian tails.

After simple calculations we obtain that
\begin{equation}\label{parffin}
{\cal Z}_N(q)\approx\frac{8}{7}C_N(q)\exp{N\left(\frac{1}{2}q^2s^2+q{\bar l}+\ln{7}\right)},
\end{equation}
where coefficient
\begin{equation}
C_N(q)=\frac{{\rm erf}\left(\frac{\ell_{\rm max}(N)-N{\bar l}-qNs^2}{\sqrt{2Ns^2}}\right)
+{\rm erf}\left(\frac{N{\bar l}+qNs^2-\ell_{\rm min}(N)}{\sqrt{2Ns^2_N}}\right)}
{{\rm erf}\left(\frac{\ell_{\rm max}(N)-N{\bar l}}{\sqrt{2Ns^2}}\right)
+{\rm erf}\left(\frac{N{\bar l}-\ell_{\rm min}(N)}{\sqrt{2Ns^2}}\right)}
\end{equation}
is responsible for surface effects which are not essential in infinite Bethe lattice;
\begin{equation}
{\rm erf}(x)=\frac{2}{\sqrt{\pi}}\int\limits^x_0{\rm e}^{-t^2}dt.
\end{equation}

Comparison of the values of $\tau(q)$ obtained numerically and calculated on the base of
Eq.~(\ref{parffin}) is presented in Fig~8. We can conclude that a theoretical formula works
in the limited interval of $q$ (small $|q|$) and satisfactory describes exponent $\tau(q)$
for relatively small $1<N<\infty$. However, we can note that there is a possibility to improve
theoretical result by inclusion of higher-order correlations $\xi^{(n)}$, that leads to the
processes beyond Markov theory.

\begin{figure}
\begin{center}
\includegraphics[width=8cm]{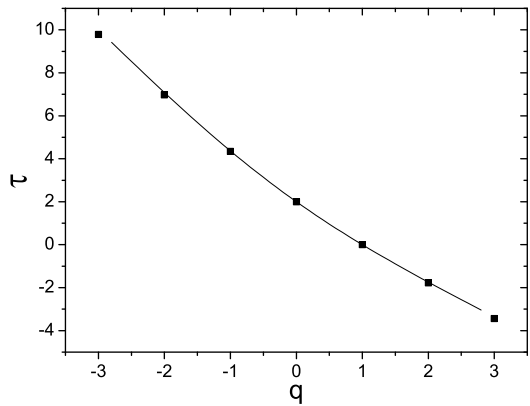}
\end{center}
\vspace*{-10mm}
\caption{\small Multifractal behavior of the exponent $\tau(q)$ for $N=5$ and $(a=0.8,\alpha=\pi/3)$.
The points and the curve are obtained on the base of partition functions (\ref{parf}),
(\ref{parffin}), respectively.}
\end{figure}

\section{Discussion}

Generally, we have concerned here with calculation of the series over Fucsian group
as group of symmetries of hyperbolic octagonal lattice with one independent complex
module. To develop an analytical approach, we have considered the model of directed
random walk on 8-branching Cayley tree as dual lattice for a given octagonal lattice.
Then, the truncated series over symmetry group was treated as the partition function
with Boltzmann weights for ensemble of lengths of tree graphs. Each of these lengths
was defined by the sum of hyperbolic distances between the tree sites visited by
directed walker and the origin.

As the result, we have analytically derived the spectrum of lengths (\ref{len})
leading to the partition function (\ref{parfM}) in the form of the Markov
multiplicative chain allowing us to apply the central limit theorem. It says about
randomizing environment, where spatial evolution is developed, and applicability of
a random walk method in our calculations. Moreover, we have explored the model within
the multifractality concept giving us a spectrum of scaling exponents.

In principal, the investigations performed might be mapped into the problem of calculation
of the Selberg trace~\cite{Hurt}, intensively used for quantization of systems on
the Riemann surfaces of genus two and requiring the knowledge of geodesics length spectrum
(see Ref.~\cite{N}, for instance). Also, in order to obtain the field observables on
Riemann surface of the fixed genus and geometry, we can ``symmetrize'' these quantities
under the action of corresponding group, that often coincides with the Dirichlet or
Poincar\'e series computation. We only note a slow convergence of these series, that
requires to use the accurate approximations which we have touched in this paper.

\section*{Appendix A. Curved Space vs Potential Field}

Let us consider a particle in an attraction potential field $U(x,y)>0$ in Euclidean
space. The Lagrangean function and the energy are
\begin{eqnarray}
L&=&\frac{1}{2}\left[\left(\frac{dx}{dt}\right)^2+\left(\frac{dy}{dt}\right)^2\right]
+\frac{1}{2}U(x,y),\label{L}\\
E&=&\frac{1}{2}\left[\left(\frac{dx}{dt}\right)^2+\left(\frac{dy}{dt}\right)^2\right]
-\frac{1}{2}U(x,y).\label{E}
\end{eqnarray}

Let potential obey the Liouville equation,
\begin{equation}\label{Liouv}
\Delta_\mathbb{E}\ln{U}=-2KU,\qquad
\Delta_\mathbb{E}=\frac{\partial^2}{\partial x^2}+\frac{\partial^2}{\partial y^2},
\end{equation}
with $K=-1$ (negative Gaussian curvature); $\Delta_\mathbb{E}$ is the Laplace
operator in Euclidean two-dimensional space.

Limiting by the case of central symmetry, when $U=U(r)$, $r=\sqrt{x^2+y^2}$,
equation (\ref{Liouv}) takes the form:
\begin{equation}\label{Liouv2}
U^{\prime\prime}+\frac{1}{r}U^{\prime}-\frac{(U^\prime)^2}{U}=2U^2,
\end{equation}
where $U^\prime=dU(r)/dr$.

Solution to Eq.~(\ref{Liouv2}) is
\begin{equation}\label{pot}
U(r)=\frac{A^2}{r^2\sinh^2(A\ln{r}+C)},
\end{equation}
where $A$, $C$ are arbitrary constants.

Following the consequence of Maupertuis variational principle, at the fixed value of
the energy $E$, the truncated action function $s$ and the time $t$ are determined from
differential relations:
\begin{eqnarray}
ds^2&=&(2E+U)(dr^2+r^2d\varphi^2),\label{metric}\\
dt^2&=&\frac{dr^2+r^2d\varphi^2}{2E+U},\label{time}
\end{eqnarray}
where $\varphi=\arctan{(y/x)}$.

Such a description of dynamical system is based on the equivalence of
the trajectories within the usual Lagrangean formalism and the geodesics
in space with metric (\ref{metric}).

In the case $E=0$, we arrive at the metric:
\begin{equation}\label{Lob}
ds^2=U(r)(dr^2+r^2d\varphi^2)
\end{equation}
with Gaussian curvature $K=-1$.

Thus we have done transition from the problem of the particle in the
potential field in Euclidean space to the problem of the free particle
in the curved Lobachevsky space. We find that the metric (\ref{Poin})
is realized when $A=1$, $C=0$.

Since the metric (\ref{Lob}) after replacement $r\to\exp{(-C/A)}r$,
$\varphi\to\varphi/A$ gives us (\ref{Poin}), these metrics are isometric.

Coming back to Eq.~(\ref{time}), relation between $t$ (time in Euclidean model)
and $s$ (proper time in Poincar\'e model now) is derived from expression:
\begin{eqnarray}
t&=&\int\frac{1}{2}(1-x^2-y^2)\sqrt{\left(\frac{dx}{ds}\right)^2+\left(\frac{dy}{ds}\right)^2}ds
\nonumber\\
&=&\frac{R^2\sqrt{1+R^2}\sinh{s}}{\sqrt{1+R^2}\cosh{s}+R}
-2R^3\arctan{\left[\left(\sqrt{1+R^2}-R\right)\tanh{\frac{s}{2}}\right]}.
\end{eqnarray}

Then one finds that trajectories for the problem (\ref{L})-(\ref{E}) can be defined in
parametric form as $\{t,x^i(t)\}=\{t(s),x^i(s)|s\in\mathbb{R}\}$, and
\begin{equation}
\frac{dx^i}{dt}=\frac{dx^i}{ds}\left(\frac{dt}{ds}\right)^{-1}.
\end{equation}

\section*{Appendix B.}

In order to describe completely the geometry of fundamental domain ${\cal F}$
in our model, it is necessary to find seven quantities,
$$
R_\pm,\ \
\phi_\pm,\ \
b,\ \
\beta,\ \
\gamma,
$$
as functions of $a$ and $\alpha$ (see Fig.~1). Parameters $R_\pm$, $\phi_\pm$ determine
the geodesics forming edges of octagon, $b$ (or $b{\rm e}^{i(\alpha+k\pi/2)}$, $k=\overline{0,3}$)
determines the location of four vertices of octagon, $\beta$ and $\gamma$ are the angles
by vertices $a{\rm e}^{ik\pi/2}$ and $b{\rm e}^{i(\alpha+k\pi/2)}$, respectively. Here
we write down seven equations giving us the values of these parameters in terms of $a$
and $\alpha$.

The first equation results from the Gauss--Bonnet theorem for surfaces of genus two:
\begin{equation}
\beta+\gamma=\frac{\pi}{2}.
\end{equation}

Next six equations are derived by using the equations of geodesics in the following
(complex) form:
\begin{equation}
\left|z-\sqrt{1+R^2_\pm}{\rm e}^{i(\phi_\pm+k\pi/2)}\right|^2=R^2_\pm,\quad
k=\overline{0,3},\quad z\in\mathbb{D}.
\end{equation}

Due to the special symmetry of octagon, it is quite enough to formulate the
necessary relations at two vertices: $z=a$ and $z=b{\rm e}^{i\alpha}$. These relations
follow from the requirement of intersection of two geodesics, labeled as $\pm$, and are
of the form:
\begin{eqnarray}
&&1+a^2-2a\sqrt{1+R^2_+}\cos{\phi_+}=0,\\
&&1+a^2-2a\sqrt{1+R^2_-}\sin{\phi_-}=0,\\
&&a^2-a\left(\sqrt{1+R^2_+}\cos{\phi_+}+\sqrt{1+R^2_-}\sin{\phi_-}\right)\nonumber\\
&&+\sqrt{1+R^2_+}\sqrt{1+R^2_-}\sin{(\phi_--\phi_+)}-R_+R_-\cos{\beta}=0,
\end{eqnarray}
and
\begin{eqnarray}
&&1+b^2-2b\sqrt{1+R^2_+}\cos{(\alpha-\phi_+)}=0,\\
&&1+b^2-2b\sqrt{1+R^2_-}\cos{(\phi_--\alpha)}=0,\\
&&b^2-b\left(\sqrt{1+R^2_+}\cos{(\alpha-\phi_+)}+\sqrt{1+R^2_-}\cos{(\phi_--\alpha)}\right)\nonumber\\
&&+\sqrt{1+R^2_+}\sqrt{1+R^2_-}\sin{(\phi_--\phi_+)}-R_+R_-\cos{\gamma}=0,
\end{eqnarray}
where $\phi_+<\alpha<\phi_-$ by construction.

As it has been noted in main text of the paper, solution to these equations is
expressions (\ref{R-phi})--(\ref{b}) derived after simple but cumbersome computations.

\section*{Appendix C.}

In this Section, we would like to discuss some properties of the kernel of multiplicative
chain (\ref{parfM}) defined at small $|q|$:
\begin{equation}
K_{i_1,i_N}(q,N)\equiv\sum\limits_{i_2=1}^8\ldots\sum\limits_{i_{N-1}=1}^8
\prod\limits_{t=2}^N {\cal P}^t(i_{t-1}|i_t).
\end{equation}

First, introducing the vector
\begin{equation}
K_{i_1}(q,N)\equiv\sum\limits_{i_N=1}^8K_{i_1,i_N}(q,N),
\end{equation}
it is worthy to note that this expression allows one the following representation:
\begin{equation}
K_i(q,N)=K_+(q,N)\delta^+_i+K_-(q,N)\delta^-_i,
\end{equation}
which is obtained on the base of semi analytical observations and reflects the symmetry
of the lattice in our model; $\delta^\pm_i$ is defined by Eq.~(\ref{delta}).
Due to this expansion, Eq.~(\ref{parfM}) is re-written as
\begin{equation}
{\cal Z}_N(q)\simeq 4\exp{(q\ell_+)}K_+(q,N)+4\exp{(q\ell_-)}K_-(q,N).
\end{equation}

Next, it is useful to define the probability of transition from initial state $i$ to
final state $f$ for $N$ steps:
\begin{equation}
P_{i,f}(q,N)\equiv\frac{K_{i,f}(q,N)}{K_i(q,N)},\qquad
\sum\limits_{f=1}^8P_{i,f}(q,N)=1.
\end{equation}
In principal, using $P_{i,f}(q,N)$, one can determine a more probable final state $f$ for
a given $q$, $N$, and initial state $i$. We can also verify that
\begin{equation}
P_{i,f}(0,N\to\infty)\to\frac{1}{8}.
\end{equation}
It means that the chain (\ref{parfM}) is ergodic at $q=0$. The same conclusion is
obtained by investigation of the mean value $\bar l$ (the mean step of graphs) which
is introduced in Eq.~(\ref{gauss}) and equal to
\begin{equation}
\bar l\equiv\lim\limits_{N\to\infty}\frac{\bar L}{N^2}\to\frac{1}{2}\bar\xi.
\end{equation}
This result is also independent on initial condition.


\end{document}